\documentclass[preprint,prd,noshowpacs]{revtex4}
\usepackage{amssymb}
\usepackage{amsmath}
\usepackage{amsfonts}
\usepackage{graphicx}
\begin{document}

\title{Inflation with a graceful exit and entrance driven by Hawking radiation}

\author{Sujoy Kumar Modak}
\email{sujoy@iucaa.ernet.in}
\affiliation{IUCAA, Post Bag 4, Ganeshkhind, Pune University Campus, Pune - 411 007, India}
\author{Douglas Singleton}
\email{dougs@csufresno.edu}
\affiliation{Physics Department, CSU Fresno, Fresno, CA 93740 USA \\
and \\
Institut f{\"u}r Mathematik, Universit{\"a}t Potsdam
Am Neuen Palais 10, D-14469 Potsdam, Germany}

\date{\today}

\begin{abstract}
We present a model for cosmological inflation which has a natural ``turn on" and a natural
``turn off" mechanism. In our model inflation is driven by the Hawking-like radiation
that occurs in Friedman-Robertson-Walker (FRW) space-time. This Hawking-like radiation
results in an effective negative pressure ``fluid" which leads to a rapid period of
expansion in the very early Universe. As the Universe expands the FRW Hawking temperature decreases
and the inflationary expansion turns off and makes a natural transition to the power law expansion of a radiation
dominated universe. The ``turn on" mechanism is more speculative, but is based on the common
hypothesis that in a quantum theory of gravity at very high temperatures/high densities 
Hawking radiation will stop. Applying this speculation to the very early Universe implies that
the Hawking-like radiation of the FRW space-time will be turned off and therefore the inflation
driven by this radiation will turn off. 
 
\end{abstract}

\maketitle

\section{Introduction}

Cosmological inflation \cite{guth} \cite{starobinsky} \cite{linde} \cite{steinhardt} 
was proposed to address the horizon problem, flatness problem and monopole problem in 
the context of Big Bang cosmology. By postulating that in the early Universe 
there was a brief period of rapid, exponential expansion one can explain, without 
fine-tuning, the observed facts that the Universe is the same in different regions which 
are causally disconnected (the horizon problem), the Universe appears to be spatially flat (the
flatness problem) and that there appears to be a much lower density of Grand Unified monopoles than
one would naively expect. However, the inflation hypothesis itself has several unanswered questions:
(i) What is the detailed mechanism for inflation? (ii) What precedes the inflationary phase or how does
inflation ``turn on"? (iii) How does the Universe make a graceful exit from this early, inflationary phase 
to standard Friedman-Robertson-Walker (FRW) radiation dominated expansion i.e. how does inflation ``turn off". 
In many of the original models \cite{guth} \cite{linde} \cite{steinhardt} inflationary expansion was driven by 
a phase transition at the Grand Unified scale. The mechanism for inflation we propose here is based on 
particle creation from the gravitational field and it need not occur at the same time/energy scale
compared to the canonical examples of inflationary mechanisms. Specifically,
we focus on particle creation connected with the Hawking-like radiation that occurs in FRW space-time.
This is similar to black hole evaporation, but time reversed. For an astrophysical size black hole Hawking
radiation is at first a very weak channel for mass/energy loss for the black hole. As the black hole
decreases in mass due to loss from Hawking radiation it gets hotter and evaporates at a faster rate.
Beyond some size Hawking radiation becomes very strong so that near the end stages of evaporation the
black hole will radiate explosively. However, near the end stages of evaporation one can no longer
trust the semi-classical calculation \cite{hawking} leading to Hawking radiation. One common speculation is
that near the end stages of evaporation where quantum gravity should become important, that Hawking
radiation will ``turn off". One concrete proposal along these lines is the suggestion that in the quantum gravity
regime space-time becomes non-commutative which leads naturally to a turning off of Hawking radiation 
in the late stages of black hole evaporation \cite{nicolini}. Applying these ideas to FRW space-time 
leads to a time reversed version of black hole evaporation. During the very earliest stages of the Universe when the
energy density is large, so that one is in the quantum gravity regime, the Hawking
radiation from the FRW would be turned off until the Universe expanded to the point when quantum gravity started to
give way to semi-classical gravity. At this point the Hawking radiation of FRW space-time would ``turn on"
and as we show below, would drive a period of exponential expansion. As the Universe expanded the Hawking
temperature of the FRW universe would decrease until the Universe becomes dominated by ordinary radiation
rather than Hawking radiation. At this point the Universe would make a gracefully transition
from inflationary expansion to the power law expansion associated with a Universe dominated by ordinary 
radiation.
  
Already in the 1930s Schr{\"o}dinger \cite{schrodinger} put forward the idea that particle creation 
can influence cosmological evolution. More recently Parker \cite{parker} and others \cite{brout}-\cite{gao}
have followed this early work of Schr{\"o}dinger with studies of how particle creation can 
affect the structure of cosmological space-times. As pointed out in \cite{prigo} there are two points
about cosmological history which are well addressed by these particle creation models. First, one can explain
very well the enormous entropy production in the early Universe via the {\it irreversible energy flow from the 
gravitational field to the created particles}. Second, since the matter creation is an {\it irreversible} 
process one avoids the initial singularity in cosmological space-times \cite{prigo}. In this model 
the Universe begins from an instability of the vacuum instead of a singularity. The Universe then rapidly 
moves through an inflationary phase followed by a radiation dominated era and finally followed by 
a matter/dust dominated era.

\section{Thermodynamics and particle creation in FRW space-time}

Our particle creation/Hawking radiation model for inflation is closely tied to thermodynamics 
in a given space-time so we begin by collecting together some thermodynamic results. The first
law of thermodynamics reads $dQ=d(\rho V)+pdV$,
where $dQ$ is the heat flow into/out of the system during some interval of cosmic time from $t$ to 
$t+dt$, $\rho$ is the energy density, $V$ is the volume and $p$ is the thermodynamic pressure. Dividing
this equation by $dt$, gives the following differential form for the first law of thermodynamics,
\begin{eqnarray}
\frac{dQ}{dt}=\frac{d}{dt}(\rho V)+p \frac{dV}{dt}. \label{dflaw}
\end{eqnarray}
For most cosmological models the assumption is made that the Universe is a  
{\it closed, adiabatic} system which means $dQ=0$. With this assumption the second law of thermodynamics, 
$dQ=TdS$, leads to a non-change in the entropy, i.e. $dS=0$, during the cosmic time interval
$dt$. This line of reasoning contradicts the observed fact that the Universe has 
an enormous entropy. This contradiction can be addressed by having irreversible particle creation 
from the gravitational field i.e. Hawking radiation from an FRW space-time. This irreversible particle
production leads to entropy production. The change in heat, $dQ$, is now completely due to 
the change of the number of particles coming from particle creation. Therefore there is a transfer of energy from 
the gravitational field to the created matter and the Universe is treated like an {\it open, adiabatic} 
thermodynamic system \cite{prigo}.

We review the relevant parts of the FRW space-time. The standard FRW metric is 
\begin{eqnarray}
ds^2=-c^2dt^2+a^2(t)\left[\frac{dr^2}{1-kr^2}+r^2(d\theta^2+\sin ^2\theta d\phi^2) \right],
\label{frw}
\end{eqnarray}
where $a(t)$ is the scale factor and $k=0, \pm 1$ is the spatial curvature of the Universe -- $k=0$ is flat, $k=-1$
is open and $k=+1$ is closed). The Einstein field equations ($G_{\mu\nu}=\frac{8\pi G}{c^4} T_{\mu\nu}$) 
for this metric have a time-time ($\mu=\nu=0$) component and space-space $\mu=\nu=i$ component
given respectively by
\begin{equation}
3\frac{\dot a^2}{a^2} +3 \frac{k c^2}{a^2} = \frac{8\pi G \rho}{c^2} ~~ , ~~~~
2\frac{\ddot{a}}{a}+\frac{\dot a^2}{a^2}+\frac{kc^2}{a^2}=-\frac{8\pi G}{c^2}p ~. \label{00}
\end{equation}
In the above equations $\rho$ is the energy density and $p$ is pressure of the matter source fluid/field. 
Combining these two equations gives the standard conservation relationship
$d(\rho V)+pdV=0$, which clearly, describes the Universe as a closed, adiabatic system with
$dQ=0$. As mentioned above this leads to $dS=0$ which then seems to contradict the very large
observed entropy of the Universe. Allowing for matter creation alters things. First in the 
presence of matter creation the equations \eqref{00} are altered. The first equation
on the left of \eqref{00} remains the same but the second equation is altered and one 
has an additional equation for the time rate of change of particle number density. These
modified and additional equations are \cite{Lima},
\begin{eqnarray}
2\frac{\ddot{a}}{a}+\frac{\dot a^2}{a^2}+\frac{kc^2}{a^2} &=& -\frac{8\pi G}{c^2}(p-p_c)~\label{feq1} \\
\frac{\dot{n}}{n}+3\frac{\dot a}{a} &=& \frac{\psi}{n}~.\label{feq2}
\end{eqnarray}
The overdot implies a time derivative, $n$ is particle number density, $ \psi$ is the 
matter creation rate and $p_c$ is the pressure due to matter creation. The matter creation rate and
the matter creation pressure are connected by the following relationship \cite{Lima},
\begin{eqnarray}
p_c=\frac{\rho + p}{3n H}\psi ~. 
\label{crp}
\end{eqnarray}
If one assumes that $\rho$ and $p$ describe a normal fluid so that one has the
energy condition $\rho + p > 0$ (assuming that $\rho >0$ this condition is known as
the weak energy condition \cite{visser}) and in addition that the matter creation rate is positive
$\psi >0$ one can see that $p_c$ of \eqref{crp} is positive and thus contributes a negative pressure to
\eqref{feq1}. Such negative pressures can drive accelerated expansion such as during the early inflationary
phase of the Universe or during the current ``dark energy" dominated era of the Universe. 
It would be economical if this negative pressure that occurs due to Hawking radiation
in FRW space-time could drive both the inflationary era {\it and} the present accelerated phase
of the Universe which is normally attributed to dark energy. We will show that while this particle
creation pressure can drive inflation it can not drive the present accelerated expansion.     

In the form \eqref{crp} one could easily explain both inflation and the current
accelerated expansion by simply choosing a matter creation rate $\psi$ to produce whatever
acceleration (if $\psi >0$) or deceleration (if $\psi <0$)  one wants. For example, if one wants
exponential expansion, $a(t) \propto e^{H t}$ one should choose $\psi = 3n H$ \cite{Lima}. However 
this choice has very little physical motivation beyond giving one the result one wanted in advance.
The strength of our proposal is that the particle creation comes from a specific mechanism -- Hawking radiation
in FRW space-time -- and as such leads to definite predictions which allow the model to be verified or ruled out. 
We will see that our mechanism does in fact lead to a particle production rate $\psi \approx 3n H$.

We now move on to a discussion of Hawking radiation and associated temperature in FRW space-time.
Since the FRW space-time is dynamical, the definition of the cosmological event horizon is subtle. However 
one can define the apparent horizon knowing the local properties of the space-time. In order to do this 
one can rewrite \eqref{frw} in the following form \cite{sp-kim}
\begin{eqnarray}
ds^2=h_{ab}dx^adx^b+\tilde r^2(d\theta^2 +\sin^2\theta d\phi^2)
\label{trmet}
\end{eqnarray}
where, $x^a=(t,r)$ and $h_{ab}=$ diag$(-c^2,a^2/(1-kr^2))$ and $\tilde r= a (t) r$. 
The position of the apparent horizon is given by the root (${\tilde r}_A$) of the equation 
$\left(h^{ab}\partial_a \tilde r\partial_b \tilde r\right)_{\tilde{r}=\tilde{r}_A}=0$. Expanding this equation over 
$t,r$ sector and simplifying we get the position of the apparent horizon ($\tilde{r}_A$) \cite{bak}
\begin{equation}
\left[h^{tt}(\partial_{t}\tilde{r})^2 + h^{rr}(\partial_r \tilde{r})^2 \right]_{\tilde{r}=\tilde{r}_A} = 0
\implies {\tilde r}_A =\frac{c}{\sqrt{H^2+\frac{kc^2}{a^2}}}. ~,\label{rah}
\end{equation}

Using the above one can find the Hawking temperature of the apparent horizon \cite{sp-kim} 
\begin{equation}
T= \frac{\hbar c \kappa}{2 \pi k_B} =  \frac{\hbar c}{2 \pi k_B} 
\left( \frac{1}{{\tilde r}_A} \right) \left| 1-\frac{{\dot {\tilde r}}_A}{2 H {\tilde r}_A} \right| . \label{htemp1}
\end{equation}
The first equality above is the standard relationship between the Hawking temperature and surface 
gravity $\kappa$ at the horizon of a given space-time. For FRW space-time the surface gravity is
$\kappa =\frac{1}{{\tilde r}_A} \left| 1-\frac{{\dot {\tilde r}}_A}{2 H {\tilde r}_A} \right|$.
Thus in general the temperature, $T$, depends on both ${\tilde r}_A$ and its
time derivative, ${\dot {\tilde r}}_A$. However, during an inflationary phase the 
Universe's scale factor takes the form $a(t) \propto \exp( {constant \times t})$ so that
$H=\frac{\dot a}{a} = constant$. If $H=constant$ satisfies $H^2 \gg c^2/a^2$ (later we show this is the case
for our model of inflation) we have from \eqref{rah} ${\tilde r}_A \approx \frac{c}{H} = constant$ and 
${\dot {\tilde r}}_A  \approx 0$.  Thus the temperature in \eqref{htemp1} simplifies to\cite{frw-hawking}
\begin{equation}
T=\frac{\hbar\sqrt{H^2+kc^2/a^2}}{2\pi k_B} \approx \frac{\hbar H }{2\pi k_B}.\label{htemp}
\end{equation}
In the final approximation we are again assuming $H^2 \gg c^2/a^2$ which as mentioned above we will
justify later. 

Before moving to a detailed calculation of how the particle creation pressure \eqref{crp}
affects the evolution of the early Universe in the case when this pressure comes
from the particle creation from Hawking radiation, we give some numerical comparisons which show that 
this mechanism is of the correct order of magnitude to explain inflation.     
Considering $H$ to be inverse of the Planck time ($t_p \approx 10^{-43} s $) gives from 
\eqref{htemp} $T \approx \frac{\hbar H}{2\pi k_B} \approx 10^{32}~ K$. On the other hand at Planck energy ($E_p$), 
gives a Planck temperature of $T_p=\frac{E_p}{k_B} \approx 10^{32}~K$. Thus the Hawking temperature
of FRW space-time at very early is around the Planck temperature. This large temperature associated with 
Hawking radiation of FRW space-time in the early Universe is a good indication that our proposed mechanism has the
proper order of magnitude to be a major factor in the early evolution of the Universe.    

Our proposed Hawking radiation mechanism for inflation is the inverse of black hole evaporate. For
astrophysical black holes the evaporation process begins very weakly -- for a black hole having
the mass of our Sun the temperature of the black body radiation emitted is $\approx 10^{-7} K$. However
at the end stages of evaporation when the black holes has a small mass the evaporation will proceed explosively.
At this point one is not justified in using the approximations that led to Hawking radiation as a thermal
spectrum and it is said that one must have in hand a quantum theory of gravity to understand these
end stages of black hole evaporation. For FRW space-time one is not justified in using Hawking radiation
results at the very early stage of the Universe; one should have a theory of quantum gravity to understand this regime.
As the Universe expands there will be a point at which the approximations leading to Hawking radiation from
FRW space-time become valid. It is at this point that our Hawking radiation from FRW space-time mechanism for
inflation ``turns on" and inflation begins. As the Universe inflates further the Hawking temperature naturally 
decreases and our inflation mechanism will automatically ``turn off". 

One can see that our proposed process is the inverse of black hole evaporation since   
the direction of radiation flux of the Hawking radiation for the apparent horizon 
in a FRW space-time is the opposite from that of a Schwarzschild black hole event horizon. For 
black holes, the created particles escape outside the event horizon towards asymptotic infinity, 
while for the apparent horizon of FRW space-time the created particles come inward from the horizon. 
Due to the isotropy of FRW space-time, the radiation is isotropic from all directions. The net result 
is an effective power gain in the Universe, given by the Stephan-Boltzmann (S-B) radiation law. 
In summary the difference in the radiation direction from a Schwarzschild black hole and from the 
FRW space-time is as follows: for black holes the time rate of energy change $P$ is negative (i.e. 
they lose power during Hawking evaporation) while for the FRW space-time the time rate of energy change,
$P$, is positive (i.e. the Universe gains energy). According to the Stephan-Boltzmann radiation law
the time rate of energy gain due to Hawking radiation is
\begin{equation}
P=+\frac{dQ}{dt}=\sigma A_H T^4, \label{power}
\end{equation}
where $\sigma=\frac{\pi^2 k_B^4}{60\hbar^3 c^2}$ is the S-B constant and $A_H$ is the area of apparent horizon.  
Now one can substitute \eqref{power} into \eqref{dflaw} but in that case the right hand side of \eqref{dflaw} 
(which is the rate of change in energy flux through the apparent horizon) has to be evaluated at $\tilde{r}= \tilde{r}_A$, so that 
\begin{equation}
\left[\frac{d}{dt}(\rho V)+p \frac{dV}{dt}\right]_{\tilde{r}=\tilde{r}_A}=\sigma A_H \left(\frac{\hbar H}{2\pi k_B}\right)^{4},
\label{genrl}
\end{equation}
where we have used \eqref{htemp}. To calculate the left hand side we first consider the volume of a sphere of arbitrary 
radius by ignoring the curvature term i.e. we take $k=0$ and the volume is given
by $V=\frac{4\pi}{3}\tilde{r}^3=\frac{4\pi}{3}r^3 a^3 (t)$. Note that here we take the radius at arbitrary
$\tilde{r}$. Only after performing the $t$ derivative in \eqref{genrl} we set $\tilde{r}=\tilde{r}_A$.   
On the other hand for the right hand side which represents the flow of energy across the apparent horizon we
take $A_H=4\pi {\tilde{r}_{A}}^2=4\pi r_A^2 a^2 (t)$ where $\tilde{r}_{A}$ comes from \eqref{rah}.
Using these expressions for the area and volume in \eqref{genrl} yields a modified continuity equation
\begin{equation}
\dot{\rho} +3 (\rho + p )\frac{\dot a}{a} = \frac{3\sigma}{c}\left(\frac{\hbar}{2\pi k_{B}}\right)^4 H^5, \label{neq}
\end{equation}
If one ignored the effect of the Hawking radiation/particle creation term on the right hand side of \eqref{neq}, by setting 
$T=0$ in \eqref{power}, then \eqref{neq} becomes $\dot{\rho} +3 (\rho + p )\frac{\dot a}{a} = 0$
which is the usual continuity equation in the absence of particle creation. 

Using $k=0$ and $H=\frac{\dot a}{a}$ we now rewrite \eqref{neq} using the first equation in \eqref{00} as 
\begin{equation}
\frac{\dot{\rho}}{\rho}+3(1+\omega)\frac{\dot a}{a} = 3 \omega_c (t) \frac{\dot a}{a} ~, \label{neq1}
\end{equation} 
where we have taken the equation of state for ordinary matter as $p = \omega \rho$ and the
time dependent equation of state due to particle creation is 
\begin{equation}
p_c(t)=\omega_c(t) \rho ~.
\label{crph}
\end{equation}
The equation of state parameter due to particle creation is
\begin{equation}
\label{omega-c} 
\omega_c (t) = \alpha \rho (t) ~~, {\rm where}  ~~\alpha = \frac{\hbar G^2}{45 c^7}= 4.8 \times 10^{-116} (J/m^3)^{-1}
\end{equation} 
The constant $\alpha$ above is essentially the inverse of the Planck energy density 
$\rho _{Planck} =\frac{c^7}{\hbar G^2} \approx 10^{114} (J/m^3)$. As we will show
later it is this constant $\alpha$ sets the time and length scale for our inflation mechanism.
This may also be different from the usual scale of inflation which is set by the Grand Unified scale. 
Moving the $\omega _c (t)$ term in \eqref{neq1} from the right hand side 
to the left hand side one can see that this particle creation term acts like
a negative pressure. For the present Universe this term is negligible. The present value of the energy density of the 
Universe is $\rho_{0}=8.91\times 10^{-10}~J/m^3$ so that $\omega _c (t_0) = \alpha \rho _0 \approx 10^{-125}$ 
term on the right hand side of \eqref{neq1} is effectively zero. Thus this effective
negative pressure can not explain the current accelerated expansion of the Universe -- 
one still needs dark energy. However in the early Universe $\rho$ can be large enough so that the particle creation pressure 
on right hand side of \eqref{neq1} dominates, and as we will see this can drive inflation and
also give a natural ``turn off" for inflation.

At this point it should be mentioned that \eqref{neq1} does not violate Wald's first axiom \cite{wald1, wald2} 
on the energy-momentum tensor which is nothing but the usual conservation equation $\nabla_{\mu} T^{\mu\nu}=0$ 
\cite{narli, paddy} . 
To see this, we note that in the absence of particle creation, the right hand side of \eqref{neq1} 
vanishes and the energy-momentum tensor has the form 
\begin{eqnarray}
T^{\mu\nu}=diag (\rho,-p,-p,-p)
\label{tgrav}
\end{eqnarray}
which satisfies the conservation equation. However in the presence of particle creation the above definition of 
$T^{\mu \nu}$ fails to simultaneously describe the conservation law and particle creation. In order to
take both features into account one needs to consider a modification $T^{\mu\nu}$ which can deal
with particle creation. Such a scenario is normally discussed in relationship to particle creation from 
black holes. Since, under the appropriate choice of vacuum state (i.e. the Unruh vacuum) 
black holes emit real particles in the form of thermal radiation, so that there is a power loss associated with 
Hawking radiation, it may appear that Wald's first axiom is violated. However, as demonstrated in \cite{elias},
for such cases it is the regularized energy-momentum tensor $\left\langle T^{\mu}_{\nu}\right\rangle$ which satisfies the 
conservation equation $\nabla_{\mu}\left\langle   T^{\mu}_{\nu}\right\rangle=0$. For the Unruh vacuum
the regularized energy-momentum tensor is
\begin{equation}
\left\langle T^{\mu \nu}\right\rangle = T^{\mu \nu}_{(gravitational)} + T^{\mu\nu}_{(boundary)} + T^{\mu\nu}_{(radiation)}.
\end{equation}
Thus it is clear that when one is dealing with particle creation it is the regularized (modified) energy-momentum 
tensor that satisfies Wald's axioms. This is exactly the picture in our case. Looking into the relation \eqref{neq1} 
one can see that the conservation equation in our case is given by $\nabla_{\mu}\tilde{T}^{\mu\nu}=0$, where the 
modified energy-momentum tensor has the form
\begin{eqnarray}
\tilde{T}^{\mu\nu} = diag (\rho, -p',-p', -p') = T^{\mu \nu}_{(gravitational)} + T^{\mu\nu}_{(radiation)} (t).
\end{eqnarray}
In the above relation $p'=p-p_c$, $T^{\mu \nu}_{(gravitational)}$ is independent of time and given by 
\eqref{tgrav} whereas the remaining part, $T^{\mu\nu}_{(radiation)} (t)$, only contains the contribution 
from $p_c(t)$ -- the particle creation pressure due to Hawking radiation.

In addition to the negative pressure \eqref{crph} associated with particle creation due 
to Hawking radiation one can also calculate the effective particle creation rate, $\psi _H$,
and compare with general result given in \eqref{crp}. Using the equation of state $p=\omega \rho$
one can re-write \eqref{crp} as 
\begin{equation}
\label{crp1}
p_c = \frac{(1+\omega )}{3 n H} \psi  \rho ~.
\end{equation}
Equating \eqref{crp1} with \eqref{crph} gives the time dependent matter creation rate 
associated with particle creation due to Hawking radiation in FRW space-time
\begin{equation}
\label{crph1}
\psi_H(t)=\frac{3n H \omega_c(t)}{(1+\omega)}.
\end{equation}
Recall that in order to have exponential expansion $a (t) \propto e^{H t}$
one needs the creation rate from \eqref{crp} to be $\psi \approx 3nH$ \cite{Lima}.
Thus from \eqref{crph1}, in order to have exponential expansion (i.e. inflation)
one needs $\omega_c(t)$ to be approximately the same size as
$1+\omega$ i.e. one needs $\omega_c(t) \approx \frac{4}{3}$ if one
assumes the equation of state for ordinary radiation i.e. $\omega =\frac{1}{3}$. Since 
$\omega_c (t) = \alpha \rho (t) $ where $\alpha = 4.8 \times 10^{-116} (J/m^3)^{-1}$, this equality
(i.e. $\omega_c(t) \approx \frac{4}{3}$) will occur when the density $\rho (t) \approx 10^{116} (J/m^3)$ 
which is approximately the Planck density. This density corresponds to the density in the early Universe.
Thus the rough calculations again point toward there being a large enough matter creation
rate, $\psi_H(t)$,  in the early Universe to drive inflationary expansion. However as the Universe expands and 
$\rho (t)$ drops the creation rate $\psi_H(t)$ will decrease and this Hawking
radiation driven mechanism for inflation will turn off. 

We now give a detailed calculation of inflation driven by Hawking radiation. Inserting 
$\omega _c (t) = \alpha \rho (t)$ into \eqref{neq1} one can integrate the resulting
equation to find the energy density $\rho$ as a function of scale factor $a$ 
\begin{equation}
\rho=\frac{Da^{-3(1+\omega)}}{1+(\frac{\alpha D}{1+\omega})a^{-3(1+\omega)}} \rightarrow 
\frac{D a^{-4}}{1+\frac{3\alpha D}{4}a^{-4}} = \frac{D}{a^4+\frac{3\alpha D}{4}}~.\label{nend}
\end{equation}
$D$ is a constant and in the last equality we have taken the equation of state of the ordinary matter
to be that of radiation (i.e. $\omega=\frac{1}{3}$) since we want the early Hawking radiation
inflation phase to be followed by a universe dominated by ordinary radiation. 
The dimensions of $D$ depend on the value of the equation of state parameter $\omega$.
Note, in the classical limit ($\hbar\rightarrow 0$), $\alpha \rightarrow 0$, the FRW Hawking radiation
effect turns off, and \eqref{nend} gives $\rho \propto a^{-3(1+\omega)}$ which is the well
known result for a universe dominated by ordinary matter with an equation of state 
$p = \omega \rho$.

There are two limits of this $\rho$ from \eqref{nend}: (i) $\alpha D \gg a^4$ so that 
$\rho \approx 4/(3 \alpha)$ and the Hawking radiation effect dominates; 
(ii) $a^4 \gg \alpha D$ so that $\rho \approx D/a^4$ which is the
energy density of an ordinary radiation dominated universe. In case (i) the energy density is constant so that
one has an effective cosmological constant which, as shown below, leads to exponential, inflationary 
expansion. In both cases (i) and (ii) the Universe is radiation dominated but for case 
(i) this means Hawking radiation of an FRW space-time and in case (ii) this means ordinary radiation. 
As one can see from the two limiting case behaviors of $\rho$ these two
types of radiation result in very different evolution.

We now want to find the time-dependence of the scale factor $a(t)$. We begin by
substituting $\rho$ (from \eqref{nend}) into the first equation in \eqref{00} to get
a differential equation for $a$ as a function of $t$ .(Recall we are
assuming that $k$ in \eqref{00} is zero or negligible compared to the other terms). It is possible 
to integrate the resulting equation for $a$ to obtain
\begin{equation}
\sqrt{\alpha D + \frac{4}{3} a^{4} }+\sqrt{\alpha D} 
\ln \left[\frac{a^2}{2\sqrt{3} \left(\sqrt{\alpha D}+ 
\sqrt{\alpha D+\frac{4}{3} a^4 }\right)}\right] = \frac{8}{3}\sqrt{\frac{2\pi G D}{c^2}}t -(K-1) \sqrt{\alpha D} ~.
\label{soln}
\end{equation}
We have written the integration constant as $-(K-1) \sqrt{\alpha D}$ where $K$ is some 
positive number greater than $1$. This will make it easier to write out
some of the later formulas. One important point to make about the scale factor, $a(t)$, in \eqref{soln} is
that it has an early exponential expansion phase (the second, logarithm term on the left hand side) 
which naturally transitions to a power law expansion (the first, power law term on the left hand side). We will discuss
these two regimes in more detail in the following subsections. That these two phases come out naturally from
the proposed inflation mechanism, without need for fine-tuning some inflaton potential, is a very attractive feature. 
 In the next following three subsections we will analyze the the early time, exponential behavior of \eqref{soln}, the
later time, power-law behavior of \eqref{soln}, and then we will discuss the possible values of $D$ and $K$.

\subsection{The very early Universe limit: $\alpha D \gg a^4$}

We first examine the limit of \eqref{soln} in the very early Universe where $a(t)$ is of a size such that
one has the limit $\alpha D \gg a^4$. In this limit \eqref{soln} becomes   
\begin{equation}
a(t) = 2 (3\alpha D)^{\frac{1}{4}} \exp \left[ {\sqrt{\frac{32\pi G}{9c^2\alpha}} ~ t}  - \frac{K}{2} \right] ~.
\label{inflation-era}
\end{equation}
Thus in this limit we find exponential expansion (inflation) with a Hubble constant given by
\begin{equation}
\label{H}
H = \frac{\dot a}{a} = \sqrt{\frac{32\pi G}{9c^2\alpha}} \approx 10^{45} \frac{1}{sec}.
\end{equation}
At this point we can return and justify some of our earlier assumption and approximations.
First, after \eqref{genrl} we assumed that $H^2 \gg k \frac{c^2}{a^2}$
is valid for $a (t)$ near the Planck size or larger (e.g. for $a \ge l_{pl} = 10^ {-35} m$).
For $a(t)$ of the Planck scale one has $\frac{c^2}{a^2} \approx 10 ^{87}$ as compared to 
$H ^2 \approx 10^{90}$ from Eq. \eqref{H}. Second, we assumed that ${\dot {\tilde r}}_A  \approx 0$. This is also 
justified since from \eqref{rah} ${{\tilde r}}_A  \approx \frac{1}{H}$ and during the inflationary 
phase $H$ is approximately constant with its value given by \eqref{H}.
Thus ${\dot {\tilde r}}_A  \approx \frac{d}{dt}(\frac{1}{H}) \approx 0$.

During inflation the standard lore is that the radius of the Universe should increase by a 
factor of $10^{26}$. Thus we need
\begin{equation}
\label{scale-a}
\frac{a(t_{end})}{a (t_{begin})} = 10^{26} = \exp \left( H \Delta t \right) ~,
\end{equation}
where $\Delta t = t_{end} - t_{begin}$, with $t_{end}, t_{begin}$ being the end and beginning time for 
this Hawking radiation driven inflation. From \eqref{H} we have $H= 10^{45}$ sec$^{-1}$ so we find that
\eqref{scale-a} gives $\Delta t \approx 6 \times 10 ^{-44}$ sec. Note that if one took the ratio in
\eqref{scale-a} to be 10 orders larger (i.e. $\frac{a(t_{end})}{a (t_{begin})} = 10^{36}$) this would yield 
$\Delta t \approx 8.3 \times 10 ^{-44}$ sec. In other words the time scale for the length of this inflation is 
set by $H$ in \eqref{H} and independent of $D$ and $K$ in  \eqref{soln}. Because $H$ in \eqref{H} is so large one does not
need a very long time, $\Delta t$, in order to inflate the Universe by many order of magnitude.

In contrast to the above mechanism of inflation, which is driven by near-Planck scale physics, the
standard picture of inflation is that it is driven by physics at the Grand Unified scale
i.e. by a Grand Unified phase transition. In this standard scenario inflation is 
thought to go from $t_{begin} \approx 10^{-36}$ sec. until $t_{end} \approx 10^{-33}$ sec. 
or $t_{end} \approx 10^{-32}$ sec. Thus for inflation driven by
a phase transition at the Grand Unified scale one has $\Delta t \approx 10^{-33} - 10^{-32}$ sec.  

\begin{figure}[h]
\centering
%\rotatebox{270}{
\includegraphics[angle=0,width=15cm,keepaspectratio]{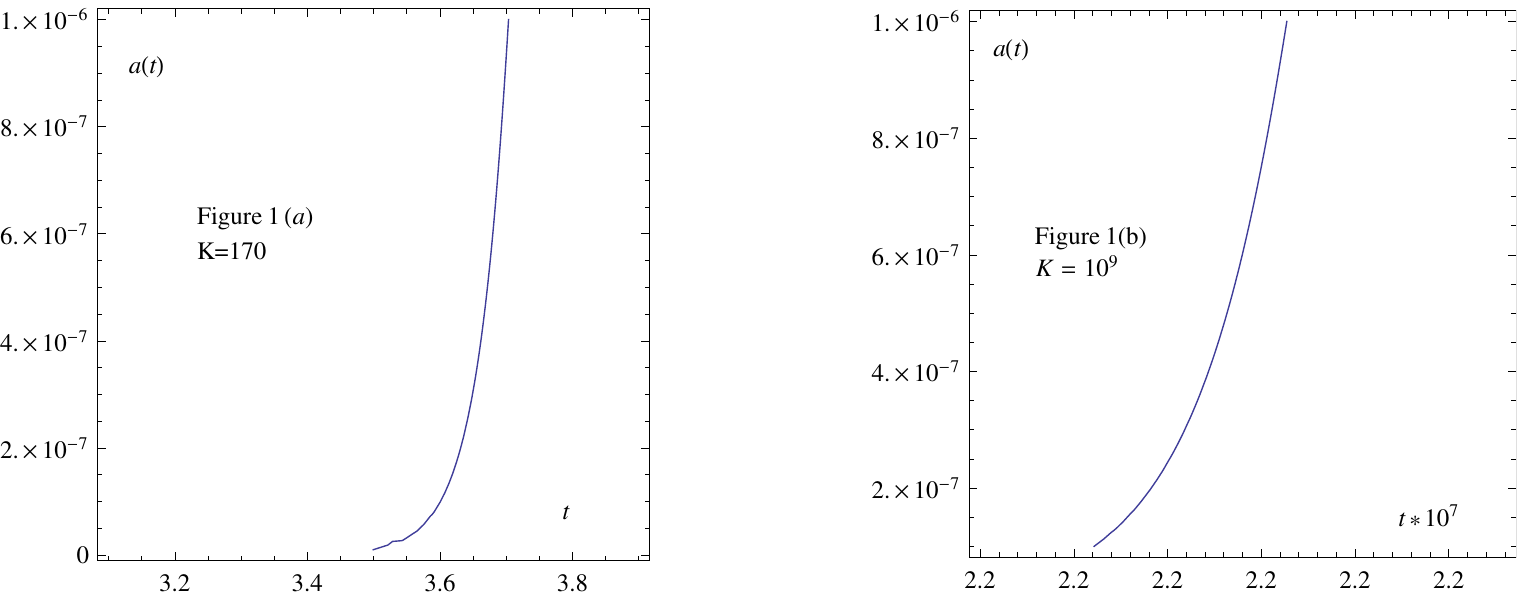}
\caption[]{\it{Scale factor $a(t)$ is plotted with respect to $t$ (in units of Planck time $t_{Pl}$) using 
equation \eqref{soln}. In (a) we fix $K=170$ and in (b) $K=10^9$. In both cases we take $D=10^{91}$ $J/m^3$. 
In this range of time $a(t)$ increases exponentially from Planck size to about $10^{-6}$ following the equation 
\eqref{inflation-era}. Because of an extremely large value of the Hubble constant \eqref{H} the lifetime of 
this inflation is very small. This is the reason why in(b) apparently time is not changing along $x$ axis. 
In fact the change in $t$ takes place after the eight decimal places and thus does not appear in the plot.}}
\label{figure1}
\end{figure}

In \eqref{figure1} we show two plots of $a vs. t$ from \eqref{soln} for the early, inflationary part of \eqref{soln}. In this 
figure we have set $D \approx 10^{91} \frac{J}{m^3}$ and two different values of $K$ are shown. This value of $D$ 
is justified in a subsequent section. From the two different values of $K$ we see that this parameter
controls when inflation starts but it does not influence how long inflation last, which in this
model is  $\Delta t \approx 10^{-43} - 10^{-44}$ sec.

\subsection{The not so early Universe limit: $a^4 \gg \alpha D$}

The scale function $a(t)$ given in \eqref{soln} will leave the regime
where the very early Universe approximation in \eqref{inflation-era} is valid, and then at some
time will reach the point where $a(t) \approx (\alpha D)^{\frac{1}{4}}$. After this intermediate
stage $a(t)$ from \eqref{soln} will continue to increase until the regime
is reached where $a^4 \gg \alpha D$. In this limit \eqref{soln} gives 
\begin{equation}
\frac{2}{\sqrt 3}a^2 + (K-1)\sqrt{\alpha D} = \frac{8}{3}\sqrt{\frac{2\pi G D}{c^2}} t ~, \label{plnew}
\end{equation}
Furthermore if the above condition is satisfied in a manner that $a^2\gg (K-1)\sqrt{\alpha D}$, one finally finds
\begin{equation}
a(t) \approx \left( \frac{32 \pi G D}{3 c^2} \right)^{1/4}  t^{1/2} ~, \label{pl}
\end{equation}
This is the usual $t^{1/2}$ power law expansion for a radiation dominated universe. 
Thus after the inflationary stage given by \eqref{inflation-era} the solution given in \eqref{soln} 
transitions into radiation dominated expansion given by \eqref{pl}.  

\begin{figure}[h]
\centering
%\rotatebox{270}{
\includegraphics[angle=0,width=15cm,keepaspectratio]{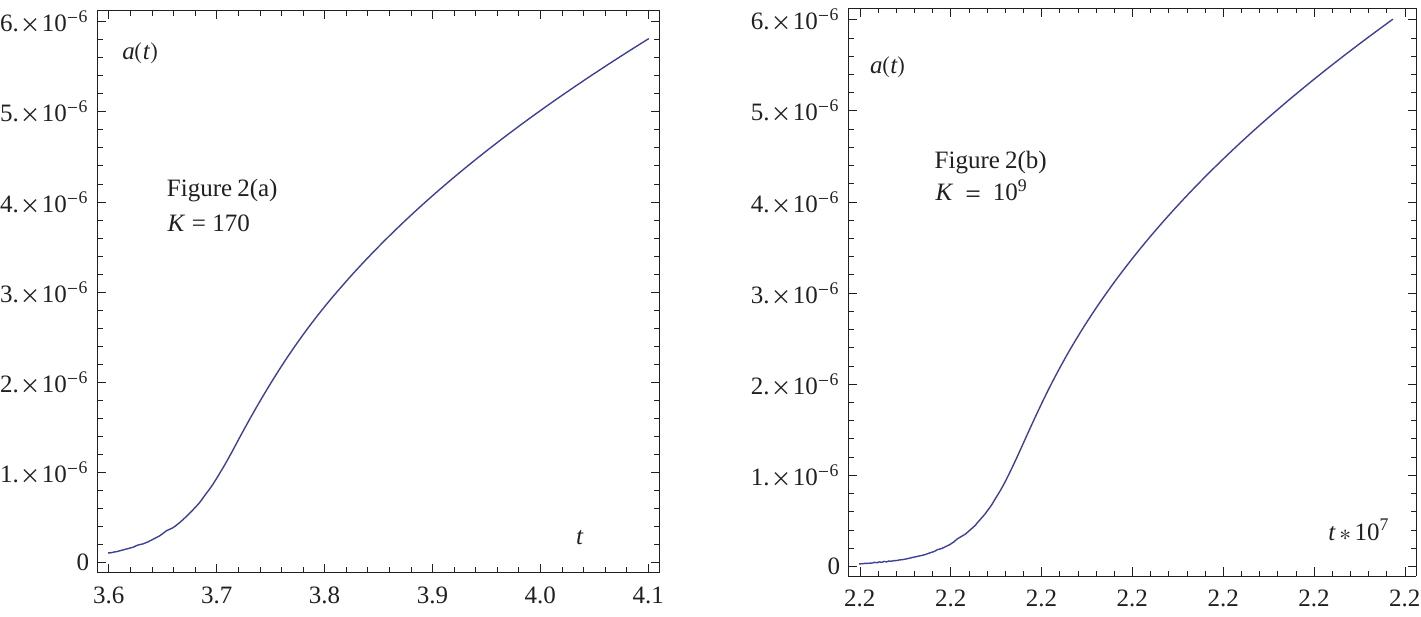}
\caption[]{\it{Scale factor $a(t)$ is plotted with respect to $t$ (in units of Planck time $t_{Pl}$)
again by using \eqref{soln}. In (a) we consider $K=170$ and in (b) $K=10^9$. As before 
we take $D=10^{91}~J/m^3$ and time changes after eight decimal places in (b). Both figures 
show that in these intermediate values of $a(t)$ the inflationary behavior \eqref{inflation-era} naturally 
makes a transition to an ordinary radiation dominated era \eqref{pl} for $a\approx 10^{-6}$. These figures 
nicely capture the end of inflation and the beginning of ordinary radiation domination.}}
\label{figure2}
\end{figure}

In \eqref{figure2} we show two plots of $a$ vs. $t$ from \eqref{soln} which shows the beginning 
of the transition from exponential inflation to $t^{1/2}$ power law inflation. Again in
this figure we have set $D \approx 10^{91} \frac{J}{m^3}$ and have the same values of $K$ 
as in \eqref{figure1}. Again the two different values of $K$ control when inflation starts but they 
do not influence its duration. 

\subsection{Determination of $D$ and $K$}

In this subsection we want to investigate possible values of the integration constants $D$
and $K$. $D$ can be set from the late time energy density of radiation. From \cite{pdg-2012}
one finds that $\Omega_{rad} =\frac{\rho_{rad}}{\rho_{crit}} \approx 4.7 \times 10^{-5}$ which
is the ratio of the radiation energy density to the critical energy density. Using the value of 
the critical energy density ($\rho _{crit} =1.7 \times 10^{-9} \frac{J}{m^3}$) we get
$\rho_{rad} = 8.0 \times 10^{-14} \frac{J}{m^3} \approx 10^{-13} \frac{J}{m^3}$ for the
present radiation energy density. Equating this with $D/a^4$ and taking $a \approx 10^{26}$  
as the present scale factor of the Universe yields $D \approx 10^{91} \frac{J}{m^3}$. Thus the
amplitude of \eqref{inflation-era} is $(\alpha D)^{\frac{1}{4}} \approx 10^{-6}$. Since for our
inflationary phase scale factor $a(t)$ given by \eqref{inflation-era} we require 
$\alpha D \gg a^4$ (which because of the $4^{th}$ power can translate 
to $(\alpha D)^{\frac{1}{4}} > a$) we see that in this picture inflation stops at a scale of 
$> 10^{-6}$ rather than $> 0.1$. However given the uncertainty in when exactly inflation ends
this is not a fatal problem. The scale of the Universe is still inflated by the same orders of magnitude --
it just starts inflating a smaller scale and ends at a smaller scale. 

Moving on to the constant $K$ one can see from Figs. 1 and 2 that this constant sets the time
scale for when inflation starts. In plots 1(a) and 2(a) where $K$ is chosen to be $K = 170$ we
find that inflation starts at $t$ a few times larger than the Planck time $t_{Pl}$. From the
figures we see that for $K=170$ inflation starts at $3.4~ t_{Pl}$ to $3.8~ t_{Pl}$. On the other hand from
plots 1(b) and 2(b) we see that for $K=10^9$ inflation starts at about $10^7 t_{Pl} \approx 10^{-36}$ sec.
This start time corresponds to the standard picture where inflation is driven by a Grand Unified
phase transition. Note that even though $K$ can shift the starting time of inflation, it can not control 
the duration which is fixed at $\Delta t \approx 10 ^{-43}$sec.

\section{Graceful entrance to inflation}

In the previous section we sketched a model for inflation driven by Hawking radiation of FRW space-time
which has a natural ``turn off" or graceful exit from inflation. We now offer speculation 
that this model of inflation driven by Hawking radiation may also have a natural ``turn on" or entrance 
to inflation. As already noted the process for inflation suggested here is the reverse of black hole 
evaporation. During post-inflation (i.e. late stage) the Hawking radiation of FRW space-time will be a weak/minor effect 
just as Hawking radiation is weak/minor effect at the beginning (i.e. early stage) of black hole evaporation.
During inflation (i.e. the very early stage) described in the section above the FRW Hawking radiation effect 
is dominate, just as during the end (i.e. very late stage) of black hole evaporation the Hawking radiation is dominate. 

During the very late stages of evaporation of a black hole there are speculations that quantum gravity effects
will turn off Hawking radiation. One particularly concrete example of this is in the non-commutative geometry
scenario \cite{nicolini} where, as the Planck scale is approached, space-time become non-commutative  
\begin{equation}
\label{nc-space-time}
[x^\mu , x^\nu]=i \theta ^{\mu \nu} ~,
\end{equation}
where $\theta ^{\mu \nu}$ is an anti-symmetric rank 2 tensor which has the dimensions of distance squared.
As a result of this non-commutativity black holes can not evaporate to arbitrarily small size, but
due to the implied uncertainty relationship between spatial coordinates -- 
$\Delta x^i \Delta x^j \ge \frac{1}{2} |\theta ^{ij}|$ for example $\Delta y \Delta z \ge \frac{1}{2} |\theta ^{yz}|$ 
-- a black hole can not shrink to zero size since then one would have $\Delta x^i =0$ in violation of 
this uncertainty relationship. Detailed analysis \cite{nicolini} shows that as a black hole 
evaporates in the non-commutative space-time characterized by \eqref{nc-space-time}           
it reaches some maximum temperature after which the black hole temperature will decrease as the
black hole continues to evaporate. At some point the Hawking temperature of the black hole goes to zero,
the evaporation process stops and one is left with a non-radiating remnant \cite{nicolini}. Applying this
picture to the FRW Hawking radiation model of inflation one would find that in the very early Universe, 
as during the late stages of black hole evaporation, the size of the Universe would be small and the FRW Hawking
temperature would be zero. Thus at this early stage there would be no inflation since the FRW Hawking radiation
would be ``turned off". The Universe would expand ``normally" according to a power law like \eqref{pl}. At some 
point the Universe would reach a size large enough not to be dominated by the uncertainty relationship coming 
from the noncommutative space-time relationship of \eqref{nc-space-time}. At this point the FRW Hawking radiation 
would ``turn on" and drive inflation until the Universe transitioned from the regime
$\alpha D \gg a^4$ to the regime $a^4 \gg \alpha D$. When the Universe entered this
regime (i.e. $a^4 \gg \alpha D$) it would undergo power law type of expansion given in \eqref{pl} 
rather than the inflationary expansion of \eqref{inflation-era}. 
 
\section{Summary}

In this paper we have proposed a mechanism for inflation based on the particle creation due to Hawing radiation
in an FRW space-time. This mechanism differs from the model of inflation driven by some
phase transition at the Grand Unified scale. This can be seen in the different time scales -- inflation driven by
a Grand Unified phase transition is thought to start at $t_{begin} \approx 10^{-36}$ sec. and last until
$t_{end} \approx 10^{-33} - 10 ^{-32}$ sec., thus having $\Delta t \approx 10^{-33} - 10 ^{-32}$ sec. Because of
the large value of $H$ in \eqref{H} (or alternatively the small value of $\alpha$ in \eqref{omega-c}) the time scale
of our proposed mechanism for inflation is  $\Delta t \approx 10^{-44} - 10 ^{-43}$ sec. which is different than the
standard time for inflation. There are two constant, $D$ and $K$, which arise in the solution of the scale 
factor \eqref{soln}. The constant $D$ is determined by matching the theoretical late-time energy density 
($\rho \approx D/a^4$) with the observed value of the present day radiation energy density 
($\rho_{rad} \approx 10^{-13} \frac{J}{m^3}$) and the present day value of $a \approx 10^{26}$. 
In this way we obtain $D \approx 10^{91} \frac{J}{m^3}$. We also get a the amplitude of the inflationary 
period expression for $a(t)$ as given in \eqref{inflation-era} namely $(\alpha D)^{1/4} \approx 10^{-6}$. This
means that this model of inflation ends when $a \le 10^{-7}$. This is six orders of magnitude smaller than the
standard picture of inflation which ends at $a \approx 0.1$. However the scale factor in our model still inflates
in size by a factor of $10^{26}$. In this picture inflation exits at a smaller scale factor than
in the canonical picture. The other constant $K$ simply shifts when inflation starts, but does not control the
duration. From Figs. 1 and 2 one can see for $K \approx {\cal O} (100)$ inflation starts near the
Planck time while for $K \approx {\cal O} (10^9)$ inflations starts near $t \approx 10^{-36}$ sec -- the
standard starting time in inflation driven by a Grand Unified phase transition. 

Because for some values of $K$ the starting time of inflation can be near the Planck time
one should worry, for these values of $K$, about the validity of the calculation of the Hawking radiation. 
For one, near the Planck scale the constants $c$, $G$ and $\hbar$ could be different from the present
day values. In particular since $\alpha$ in \eqref{omega-c} -- and therefore $H$ in \eqref{H} -- 
depend on $c$ to the seventh power, having a different value of $c$ at these early, near-Planck times
by even one order of magnitude would greatly change the scale of Hawking radiation driven inflation mechanism proposed here. 
If $c$ were one order of magnitude smaller in these very early times the energy scale of the Hawking radiation driven
inflation would shift to be more in line with that of the Grand Unified phase transition mechanism for inflation.
In this paper we simply stick to the simplest assumption -- that $c$, $G$ and $\hbar$ -- have the present, constant
values even at these early, near-Planck times. We hope later to investigate the possibility that $c$, $G$ and/or $\hbar$ 
have different value at these early times. 

In this picture of Hawking radiation driven inflation the time scale is set by $\alpha$ in \eqref{omega-c}. Setting
aside this definite scale prediction for a moment - allowing for an arbitrary scale $\alpha$ - we note that one 
might regard \eqref{nend}, and the resulting scale factor $a(t)$ in \eqref{soln}, as a good phenomenological 
model for the time development of the size of the Universe which naturally includes exponential expansion with 
power law expansion in a single expression. 

The inflation mechanism presented here is the time reversal of black hole evaporation. For a black hole 
in the {\it early stages} of evaporation via Hawking radiation, the radiation is a weak effect, 
barely changing the mass and space-time of the black hole;
for an FRW universe in its {\it late stages} the Hawking radiation is a weak effect having effectively no effect on
the expansion rate of the Universe. For a black hole in the {\it late stages} of evaporation via Hawking radiation, the
radiation is a dominant effect, which plays a significant role in the change of the black hole's mass and the structure of
the space-time; for an FRW universe in its {\it early stages} the Hawking radiation is a huge effect and leads to an
enormous expansion rate \eqref{H} for the Universe. In the {\it very late stages} of black hole evaporation it is
postulated that quantum gravity effects will shut off Hawking radiation; for an FRW space-time we postulate that
in the {\it very early stages} quantum gravity effects will shut off Hawking radiation and the associated exponential 
expansion \eqref{inflation-era}.     

There have been other works that have studied the role of particle creation in
the evolution of the Universe \cite{schrodinger} -- \cite{Lima}. The present proposal is similar to the 
work of \cite{prigo} which views particle creation as an irreversible process from energy transfer and entropy
production from the gravitational field to the particles. The difference in the present work is that 
we have proposed a very specific particle creation mechanism namely the Hawking radiation associated with FRW 
space-time. The FRW Hawking radiation gives rise to an effective negative pressure
evolution equation for the energy density, $\rho$, \eqref{neq} \eqref{neq1}. 
The resulting $\rho$ given in \eqref{nend} leads to a time dependent scale factor
$a(t)$ given in \eqref{soln} which has two regimes -- one where $\alpha D \gg a^4$ with the
resulting $a(t)$ being exponential/inflationary expansion as given in \eqref{inflation-era} and one 
where $a^4 \gg \alpha D$ with the resulting $a(t)$ being power law expansion as given in \eqref{pl}. There is
a natural transition from inflationary expansion to power law expansion so that this model for inflation 
has a graceful exit from inflationary behavior. Finally based on the inverse similarity between black hole
evaporation and this FRW Hawking radiation model of the evolution of the scale factor $a(t)$, where the
period of late time black hole evaporation corresponds to early period of the Universe (and visa versa),
we have given some speculation as to how FRW Hawking radiation mechanism  for inflation may ``turn on" due
to non-commutative space-time effects. Thus the FRW Hawking radiation picture for the evolution of $a(t)$
provides not only a graceful exit to inflation as well as a possible graceful entrance.  
 
One final comment - this inflation mechanism has a feedback mechanism which forces the scale factor, $a(t)$, to be uniform.
For example, if one assumed that the scale factor also had a dependence on $r$ (i.e. $a(r,t)$) the  Hawking radiation
inflation mechanism would tend to erase this $r$ dependence. If, $a(r,t)$ were smaller for some $r$ this would imply a
higher Hawking temperature and more rapid expansion. This would push those regions of $r$ with smaller scale factor, $a$, 
to expand more rapidly until they were the same as the scale factor in other regions.  
If, $a(r,t)$ were larger for some $r$ this would imply a lower Hawking temperature and less rapid expansion. 
This would push those regions of $r$ with larger scale factor, $a$, to expand less rapidly until they were the 
same as the scale factor in other regions.

{\bf Acknowledgment:} DS is supported by a DAAD grant.

\end{document}